\documentclass[final,nonatbib]{article}
\usepackage[numbers]{natbib}
\usepackage[final]{neurips_2024_ml4ps}
\usepackage{macros_aas}
\usepackage[utf8]{inputenc} 
\usepackage[T1]{fontenc}    
\usepackage{hyperref}       
\usepackage{url}            
\usepackage{booktabs}       
\usepackage{amsfonts}       
\usepackage{nicefrac}       
\usepackage{microtype}      
\usepackage[dvipsnames]{xcolor}         
\usepackage{microtype}
\usepackage{graphicx}
\usepackage{subfigure}
\usepackage{booktabs} 
\hypersetup{
	colorlinks   = true, 
    urlcolor     = blue, 
    linkcolor    = blue, 
    citecolor    = red   
 }
\usepackage{amsmath}
\usepackage{amssymb}
\usepackage{mathtools}
\usepackage{amsthm}
\usepackage[capitalize,noabbrev]{cleveref}
\usepackage{algorithm}
\usepackage{algorithmic}
\usepackage[textsize=tiny]{todonotes}
\usepackage{typed-checklist}
\usepackage{ifthen}

\newcommand{\torch}{{\tt Torch}}
\newcommand{\torchvision}{{\tt Torchvision}}
\newcommand{\numpy}{{\tt Numpy}}
\newcommand{\scipy}{{\tt Scipy}}
\newcommand{\pytorch}{{\tt PyTorch}}
\newcommand{\matplotlib}{{\tt Matplotlib}}

\newcommand{\deeplenstronomy}{{\tt deeplenstronomy}}
\newcommand{\lenstronomy}{{\tt lenstronomy}}

\newcommand{\seaborn}{{\tt Seaborn}}
\newcommand{\sklearn}{{\tt Sklearn}}

\newcommand{\astropy}{{\tt Astropy}}

\newcommand{\pandas}{{\tt Pandas}}

\newcommand{\python}{{\tt Python}}

\newcommand{\myfig}{Fig.}

\newcommand{\mytable}{Table}
\newcommand{\myappendix}{Appendix}
\newcommand{\mysection}{\S}

\newcommand{\modelmveuda}{MVE-UDA}
\newcommand{\modelmve}{MVE-only}
\newcommand{\datasource}{source}
\newcommand{\datatarget}{target}
\newcommand{\sersic}{S\'ersic}

\newcommand{\einsteinrad}{\theta_{\mathrm{E}}}
\newcommand{\eccentricity}[2]{e_{\mathrm{#1},#2}}

\newcommand{\myarcsec}{$^{\prime\prime}$}

\newcommand{\lossuda}{L_{\mathrm{UDA}}}
\newcommand{\lossmve}{L_{\mathrm{MVE}}}
\newcommand{\losstot}{L_{\mathrm{Tot}}}
\newcommand{\lossfactoruda}{\alpha_{\mathrm{UDA}}}
\newcommand{\betanll}{\beta_{\mathrm{NLL}}}
\newcommand{\nll}{\mathcal{L}_{\beta-\mathrm{NLL}}}

\newcommand{\residualbasicdefinition}{\einsteinrad{}_{, \mathrm{pred}} -\einsteinrad{}_{, \mathrm{true}}}

\newcommand{\sigmaal}{\sigma_{\mathrm{al}}}
\newcommand{\sigmaalmeanbin}{\bar{\sigma}_{\mathrm{al}}}

\newcommand{\residualmedian}{\residualmean_{\mathrm{med}}}
\newcommand{\sigmaalmedian}{\sigmaalmean_{\mathrm{med}}}
\newcommand{\residualmean}{\langle\delta\einsteinrad\rangle}
\newcommand{\sigmaalmean}{\langle\sigmaal\rangle}
\newcommand{\correlation}{R^2}
\newcommand{\correlationmean}{\langle\correlation\rangle}
\newcommand{\nllmean}{\langle\nll\rangle}

\theoremstyle{plain}

\theoremstyle{definition}

\theoremstyle{remark}

\newcommand{\authoraffil}[2]{$^{#1}$ \\ #2} 
\newcommand{\assignaffilnumber}[2]{$^#1$#2} 

\newcommand{\nord}{Brian D. Nord}
\newcommand{\ciprijanovic}{Aleksandra \'{C}iprijanovi\'{c}}

\newcommand{\agarwal}{Shrihan Agarwal}

\newcommand{\nordemail}{\url{nord@fnal.gov}}
\newcommand{\ciprijanovicemail}{\url{aleksand@fnal.gov}}

\newcommand{\agarwalemail}{\url{shrihan@uchicago.edu}}

\newcommand{\uchicagoAA}{Department of Astronomy and Astrophysics, University of Chicago, Chicago, IL 60637}
\newcommand{\kicp}{Kavli Institute for Cosmological Physics, University of Chicago, Chicago, IL 60637}
\newcommand{\fermilab}{Fermi National Accelerator Laboratory, Batavia, IL 60510}

\newcommand{\ackfundingdeepskies}{We acknowledge the Deep Skies Lab as a community of multi-domain experts and collaborators who’ve facilitated an environment of open discussion, idea generation, and collaboration. This community was important for the development of this project.\\ \\ }

\newcommand{\ackfundingdoegeneral}{Work supported by the Fermi National Accelerator Laboratory, managed and operated by Fermi Research Alliance, LLC under Contract No. DE-AC02-07CH11359 with the U.S. Department of Energy. The U.S. Government retains and the publisher, by accepting the article for publication, acknowledges that the U.S. Government retains a non-exclusive, paid-up, irrevocable, world-wide license to publish or reproduce the published form of this manuscript, or allow others to do so, for U.S. Government purposes.\\ \\ }

\newcommand{\ackfundingdoeecanord}{This material is based upon work supported by the Department of Energy under grant No. FNAL 21-25.\\ \\ }

\newcommand{\contribagarwal}{
Agarwal: Methodology, Formal analysis, Software, Validation, Data Curation, Investigation, Writing - Original Draft
\\ \\}

\newcommand{\contribciprijanovic}{
\'{C}iprijanovi\'{c}: Conceptualization, Methodology, Formal analysis, Writing - Review \& Editing, Supervision, Project administration
\\ \\}

\newcommand{\contribnord}{
Nord: Conceptualization, Methodology, Formal analysis, Resources, Writing - Original Draft, Writing - Review \& Editing, Supervision, Project administration, Funding acquisition
\\ \\}

\author{
\agarwal\authoraffil{1}{\agarwalemail} \And
\ciprijanovic\authoraffil{1,2}{\ciprijanovicemail} \And
\nord\authoraffil{1,2,3}{\nordemail} \\ \\
\assignaffilnumber{1}{\uchicagoAA}\\
\assignaffilnumber{2}{\fermilab}\\
\assignaffilnumber{3}{\kicp}
}
\newcommand{\figureresidual}[2]{
\begin{figure*}[htp!]
    \centering
    \includegraphics[scale=1]{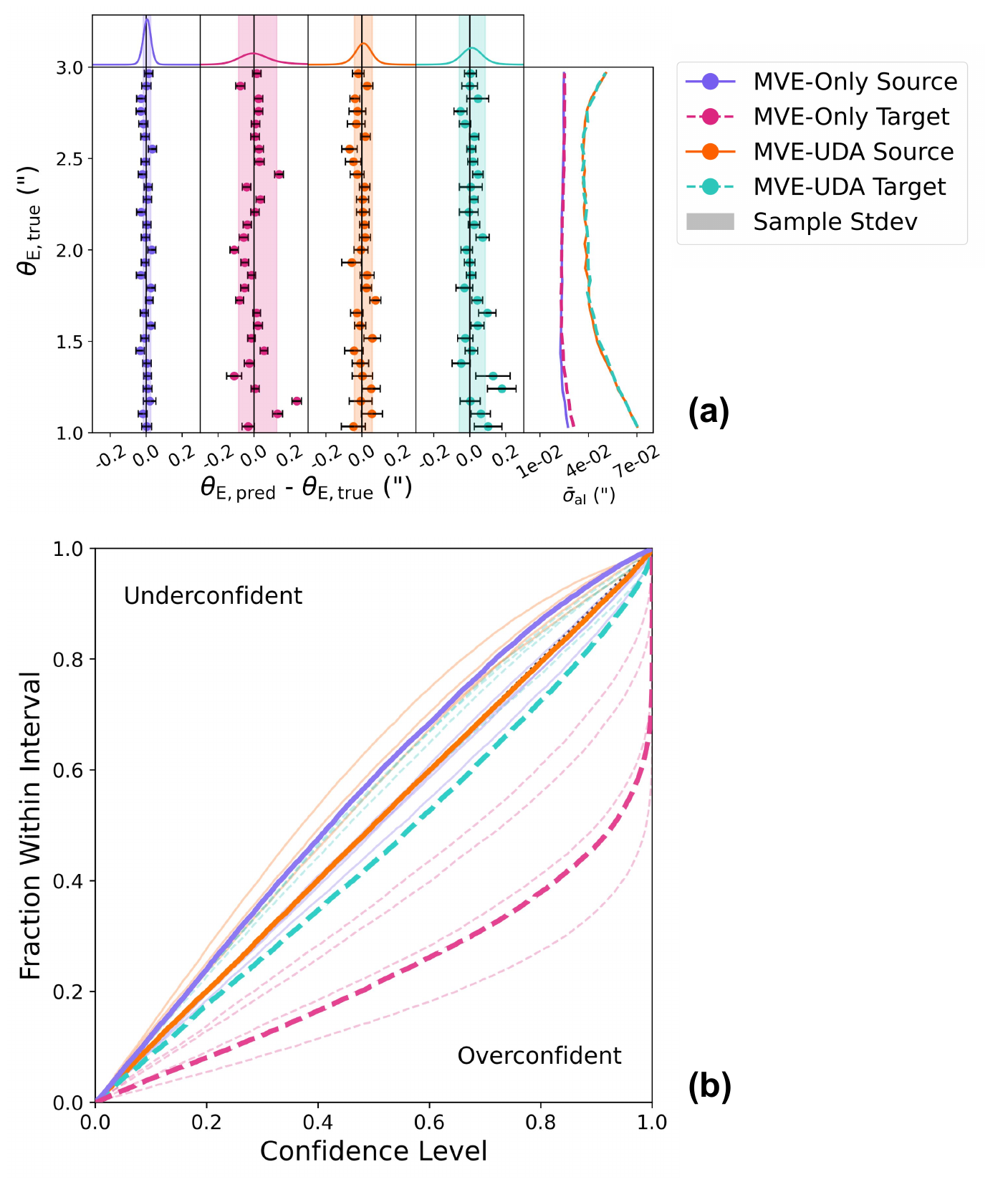}
    \vspace{-0.73cm}
    \caption{#1} #2
\end{figure*}
}

\newcommand{\figurelatent}[2]{
\begin{figure*}[htp!]
    \centering
    \includegraphics[scale=1]{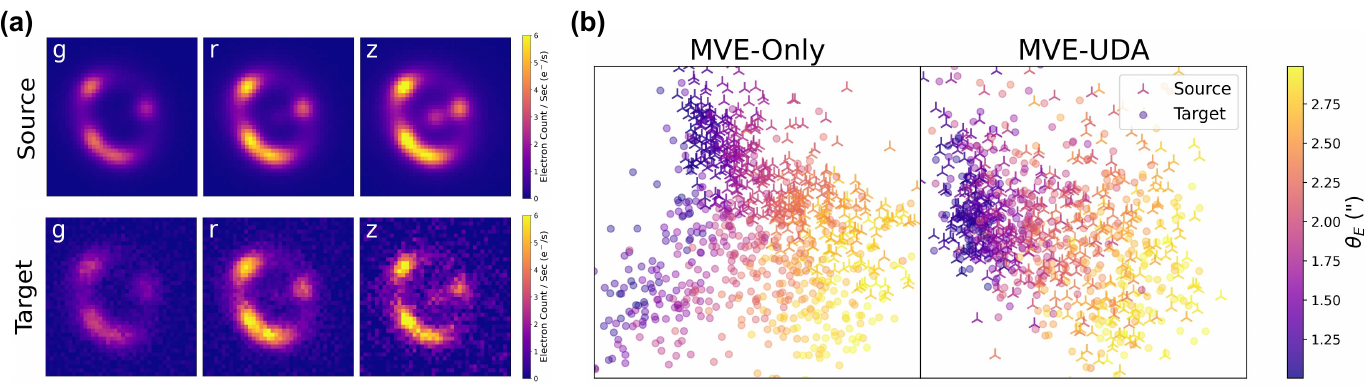}
      \vspace{-0.32cm}
    \caption{#1} #2
\end{figure*}
}

\newcommand{\tableresultsmodel}[2]
{
{\renewcommand{\arraystretch}{1.1}%
\begin{table}
   \centering
   \noindent\begin{minipage}[b]{0.99\columnwidth}
   \centering
    \caption{#1} #2
  \centering
  \fontsize{9}{9}\selectfont
  \setlength{\tabcolsep}{2.8pt}
  \begin{tabular}{ l |  c c  c c | c c  c c } 
    \toprule
                        & \multicolumn{4}{c|}{Selected}                                                       & \multicolumn{4}{c}{Median}                              \\
    \midrule
                        & \multicolumn{2}{c}{\textbf{(a)} Residual $\residualmean{}$} & \multicolumn{2}{c|}{\textbf{(b)} Uncertainty $\sigmaalmean{}$} & \multicolumn{2}{c}{\textbf{(c)} Residual $\residualmedian{}$}     & \multicolumn{2}{c}{ \textbf{(d)} Uncertainty $\sigmaalmedian{}$ }\\         
    \midrule
    Model               & Source     & Target                       & Source    & Target                            & Source    & Target                            & Source    & Target                \\ 
    \toprule
    \modelmve           & $0.0201$   & $0.0818$                     & $0.0243$  & $0.0253$                          & $0.0164$  & $0.0585$                          & $0.0203$  & $0.0205$              \\ 
    \modelmveuda        & $0.0358$   & $0.0425$                     & $0.0489$  & $0.0503$                          & $0.0389$  & $0.0461$                          & $0.0628$  & $0.0628$              \\ 
    \toprule
\end{tabular}
\end{minipage}
\end{table}
}
}

\newcommand{\tableresultsmodelappendix}[2]
{
{\renewcommand{\arraystretch}{1.1}%
\begin{table}
   \centering
   \noindent\begin{minipage}[b]{0.99\columnwidth}
   \centering
    \caption{#1} #2
  \centering
  \begin{tabular}{ l | c | c c  c c } 
    \toprule
                      &   & \multicolumn{2}{c}{\modelmve}       & \multicolumn{2}{c}{\modelmveuda}                             \\         
    \midrule
    Metric               & Seed & Source            & Target            & Source                    & Target  \\ 
    \toprule
    Residual: $\residualmean{}$	& 56 & $0.0164$ & $0.0693$ & $0.0358$ & $0.0436$ \\
		& 11 & $0.0149$ & $0.0287$ & $0.0389$ & $0.0425$ \\
		& 31 & $0.0201$ & $0.0585$ & $0.0386$ & $0.0461$ \\
		& 6 & $0.0150$ & $0.0818$ & $0.0484$ & $0.0510$ \\
		& 63 & $0.0174$ & $0.0240$ & $0.0452$ & $0.0551$ \\
    \midrule
    Uncertainty: $\sigmaalmean{}$& 56 & $0.0243$ & $0.0253$ & $0.0489$ & $0.0503$ \\
		& 11 & $0.0180$ & $0.0179$ & $0.0602$ & $0.0599$ \\
		& 31 & $0.0269$ & $0.0239$ & $0.0634$ & $0.0634$ \\
		& 6 & $0.0192$ & $0.0199$ & $0.0678$ & $0.0678$ \\
		& 63 & $0.0203$ & $0.0205$ & $0.0628$ & $0.0628$ \\
  \midrule
Correlation: $\correlationmean{}$	& 56 & $0.9986$ & $0.9642$ & $0.9924$ & $0.9835$ \\
		& 11 & $0.9988$ & $0.9939$ & $0.9917$ & $0.9897$ \\
		& 31 & $0.9979$ & $0.9727$ & $0.9922$ & $0.9886$ \\
		& 6 & $0.9988$ & $0.9418$ & $0.9880$ & $0.9861$ \\
		& 63 & $0.9984$ & $0.9968$ & $0.9889$ & $0.9832$ \\
  \midrule
  NLL Loss: $\nllmean{}$	& 56 & $-3.3603$ & $4.5586$ & $-2.6600$ & $-2.4204$ \\
		& 11 & $-3.4737$ & $-1.0705$ & $-2.5098$ & $-2.4385$ \\
		& 31 & $-3.1443$ & $15503.4180$ & $-2.4316$ & $-2.2854$ \\
		& 6 & $-3.4925$ & $25.4278$ & $-2.2687$ & $-2.2070$ \\
		& 63 & $-3.2745$ & $-2.6643$ & $-2.2982$ & $-2.0623$ \\
  \toprule
\end{tabular}
\end{minipage}
\end{table}
}
}

\newcommand{\tablenetworkmve}[2]{
\begin{table}
  \centering
  \noindent\begin{minipage}[b]{\columnwidth}
  \centering
    \caption{#1} #2
  \centering
  \begin{tabular}{l l c}
    \hline   
        Layer       & Output shape      & Parameters \\
    \hline \hline
        Conv2d      & [-1, 8, 40, 40]   & $k=3$, $s=1$\\ 
    \midrule
        BatchNorm2d & [-1, 8, 40, 40]   & $k=3$, $s=1$\\ 
    \midrule
        MaxPool2d   & [-1, 8, 20, 20]   & $k=2$, $s=2$  \\
    \midrule
        Conv2d      & [-1, 16, 20, 20]  & $k=3$, $s=1$\\ 
    \midrule
        BatchNorm2d & [-1, 16, 20, 20]  & $k=3$, $s=1$\\ 
    \midrule
        MaxPool2d   & [-1, 16, 20, 20]  & $k=2$, $s=2$  \\
    \midrule
        Conv2d      & [-1, 32, 10, 10]  & $k=3$, $s=1$\\ 
    \midrule
        BatchNorm2d & [-1, 32, 10, 10]  & $k=3$, $s=1$\\ 
    \midrule
        MaxPool2d   & [-1, 32, 5, 5]    & $k=2$, $s=2$  \\
    \midrule
        Linear      & [-1, 128]         & - \\
    \midrule
        Linear      & [-1, 32]          & - \\
    \midrule
        Linear      & [-1, 2]           & - \\
  \hline
\end{tabular}
\end{minipage}
\end{table}
}

\newcommand{\tablesetup}[2]
{
{\renewcommand{\arraystretch}{1.05}
\begin{table}
   \centering
   \noindent\begin{minipage}[b]{0.99\columnwidth}
   \centering
    \caption{#1} #2
  
  \centering
  \fontsize{9}{9}\selectfont
  \setlength{\tabcolsep}{2.7pt}
  \begin{tabular}{ l l c}
  \toprule   
 \multicolumn{2}{c}{Parameter}                    & Prior                \\ 
  \toprule
  \multicolumn{3}{c}{Lens light profile}\\
\hline
 Einstein radius   &  $\einsteinrad$ (\myarcsec)& $\mathcal{U}(1.0,3.0)$    \\ 
\sersic{} index &   $n$     & $\mathcal{U}(2.0,5.0)$ \\
 Scale radius & $R$ (\myarcsec)     & $\mathcal{U}(1.0,2.5)$ \\ 
Eccentricity &$\{e_\mathrm{l,1}, e_\mathrm{l,2}\}$     & $\mathcal{U}(-0.2,0.2)$ \\
External shear &  $\{\gamma_{1},\gamma_{2}\}$     & $\mathcal{U}(-0.05,0.5)$ \\ 
    \hline
\multicolumn{3}{c}{Source light profile}\\
    \hline
\sersic{} index & $n$     & $\mathcal{U}(2.0,4.0)$ \\
Scale radius &  $R$  (\myarcsec)    & $\mathcal{U}(0.5,1.0)$ \\ 
Eccentricity & $\{e_\mathrm{s,1},e_\mathrm{s,2}\}$     & $\mathcal{U}(-0.2,0.2)$ \\
Relative angular positions &    $\{x, y\}$ (\myarcsec) & $\mathcal{U}(-0.5,0.5)$ \\ 
  \toprule
  
\end{tabular}
\end{minipage}
\end{table}
}
}

\title{
Neural network prediction of strong lensing systems with domain adaptation and uncertainty quantification
}

\begin{document}
\author{
\agarwal\authoraffil{1}{\agarwalemail} \And
\ciprijanovic\authoraffil{1,2}{\ciprijanovicemail} \And
\nord\authoraffil{1,2,3}{\nordemail} \\ \\
\assignaffilnumber{1}{\uchicagoAA}\\
\assignaffilnumber{2}{\fermilab}\\
\assignaffilnumber{3}{\kicp}
}
\maketitle


\begin{abstract}
Modeling strong gravitational lenses is computationally expensive for the complex data from modern and next-generation cosmic surveys.
Deep learning has emerged as a promising approach for finding lenses and predicting lensing parameters, such as the Einstein radius.
Mean-variance Estimators (MVEs) are a common approach for obtaining aleatoric (data) uncertainties from a neural network prediction.
However, neural networks have not been demonstrated to perform well on out-of-domain target data successfully --- e.g., when trained on simulated data and applied to real, observational data. 
In this work, we perform the first study of the efficacy of MVEs in combination with unsupervised domain adaptation (UDA) on strong lensing data.
The \datasource{} domain data is noiseless, and the \datatarget{} domain data has noise mimicking modern cosmology surveys.
We find that adding UDA to MVE increases the accuracy on the \datatarget{} data by a factor of about two over an MVE model without UDA.
Including UDA also permits much more well-calibrated aleatoric uncertainty predictions.
Advancements in this approach may enable future applications of MVE models to real observational data.
\end{abstract}

\section{Introduction and Related Work}
\label{sec:introduction}
Strong gravitational lensing provides critical insights into galaxy evolution, dark matter, and dark energy~
\cite{albrecht_report_2006, 2023arXiv230705714T, 2010ARA&A..48...87T, 2004PhRvD..70d3534L, 2023MNRAS.524.6159K, 2024MNRAS.tmp.1779G, 2024ApJ...970..143T, 2015salt.confE..16S}.
Modern cosmological surveys~\cite{the_dark_energy_survey_collaboration_dark_2005, flaugher_dark_2015, 2022PASJ...74..247A, 2023PhRvD.108l3520M, 2013ExA....35...25D, 2019A&A...625A...2K, ivezic_lsst_2019, 2024arXiv240513491E, schaerer_first_2022, brummel-smith_inferred_2023, eifler_cosmology_2021, la_plante_prospects_2023, wang_high_2022-1}
are expected to contain 10$^3$ - 10$^5$ lensing systems~\cite{2016ApJ...827...51N, 2024arXiv240608919S, Collett2015}.
Traditional lens finding techniques have relied heavily on human-intensive image reviewing~\cite{2022ApJS..259...27O, 2020MNRAS.494.1308N, 2022A&A...668A..73R, 2023MNRAS.523.4413R, 2016MNRAS.455.1171M}, and modeling has relied on computationally-intensive analytic likelihood-fitting~\cite{Birrer_2015, Lefor_2013, 2010GReGr..42.2151K, 2023ApJ...953..189D, 2022MNRAS.517.3275E}.
This has motivated supervised deep learning-based techniques like neural network classification and regression to be applied to strong lensing in addition to a wide variety of cosmology topics~\cite{2019A&A...625A.119M, 2022ApJ...932..107S, 2017Natur.548..555H}.
Obtaining uncertainties is important for these areas of study~\cite{levasseur_uncertainties_2017}. 
They can be obtained in network regression through a variety of methods --- e.g.,  MC Dropout~\cite{2015arXiv150602142G, hasan_controlled_2022, 2020arXiv200802627V, 2021arXiv211004286L}, Bayesian Neural Networks~\cite[BNNs;][]{2020arXiv200601490C, 2021arXiv210613594C, 2023arXiv230916314A, 2020arXiv200706823V, 2020arXiv200612024G}, mean-variance estimation~\cite{2023arXiv230208875S, 2023npjCM...9..225T, 2019arXiv190603260D, stoppa_autosourceid-featureextractor_2023}, Deep Ensembles~\cite{carrete_deep_2023, gawlikowski_survey_2023, stoppa_autosourceid-featureextractor_2023, 2016arXiv161201474L, 2021arXiv211013511E, 2021arXiv210402395G, 2022arXiv220206985A}, Deep Evidential Regression~\cite{2024arXiv240101484Y, 2021arXiv210406135M, 2022arXiv220510060M, 2019arXiv191002600A}, and Simulation-Based Inference~\cite{CB2019, Legin2021, Legin2023, WagnerCarena2022}.
Once trained, these methods are typically very fast compared to traditional parametric modeling methods~\cite{levasseur_uncertainties_2017}.
However, all of these models face the challenge that there is insufficient observational data for training and instead rely on realistic simulations ~\cite{lenstronomy:2018, deeplenstronomy, 2006MNRAS.367.1367A, 2016ApJ...828...54L, 2019MNRAS.482.2823P, 2021MNRAS.501.4657R, 2020Symm...12..494P}.

Despite the realism, simulated data can differ from real, observational data --- e.g., the image noise parameters, the range of astrophysics parameters, or the range of cosmology parameters.
Sometimes, real data is used directly in training \cite{ huang_finding_2020, huang_discovering_2021} or is combined with simulated data \cite{zaborowski2023}.
The differences between the training data (\datasource{} domain data) and the real observational data (\datatarget{} domain data) constitute domain shifts between data distributions that cause models to favor the \datasource{} (training) data~\cite{2024arXiv240212627L, 2024arXiv240302714H, 2021arXiv210302503Z}.
Typically, this problem arises when there are few or no labels for the \datatarget{} data for the model to train on~\cite{wilson_survey_2020, 2019arXiv190105335K}. 
Domain adaptation (DA) is a class of deep learning techniques that help neural networks adapt to domain shifts so that the feature spaces of the source and target data domains align when the domain-adapted model is applied~\cite{2020arXiv201003978F, 2023arXiv230203133H, 2023arXiv230903879E, motiian_unified_2017, zhang_transfer_2020, zhuang_comprehensive_2020}.
Unsupervised domain adaptation (UDA) is a subclass of techniques that use unlabeled target data~\cite{2023arXiv230100265F, 2023arXiv230302302X, 2022arXiv220807422L}.
Studies have explored DA in many fields, including cosmology and strong lensing~\cite{roncoli_domain_2024, ciprijanovic_semi-supervised_2022, triess_survey_2021, ciprijanovic_domain_2020, ciprijanovic_deepmerge_2021, ciprijanovicRobustnessDeepLearning2021, ciprijanovic_deepadversaries_2022, ciprijanovicDeepAstroUDASemiSupervisedUniversal2023, 2023arXiv230500002A, 2024arXiv240402973W}.    
In this work, we combine MVE and UDA and compare the performance of \modelmveuda{} and \modelmve{} models on strong lensing data in two domains that are distinguished by the noise in the images.

\section{Methods: Lensing, Mean-variance Networks, and Domain Adaptation}
\label{sec:methods}

\textbf{Physics of strong lensing:} 
Galaxy-scale strong lensing occurs when a foreground lens galaxy deflects light from a background galaxy, creating a magnified and warped image of the background object.
This distorted image is the primary observable (\myfig{}~\ref{fig:latent}(a)) for predicting physics parameters.
Multiple kinds of noise sources ---  e.g., atmosphere, sky brightness, CCD readout, and photon counting --- can further distort the image.
The Einstein radius $\einsteinrad{}$ indicates the spatial scale of the lensing system and depends on the lens galaxy mass distribution, which is complex but can often be modeled with a 5-parameter singular isothermal ellipsoid (SIE), including the Einstein radius~\cite{Narayan1996}. 
We predict $\einsteinrad{}$. 

\textbf{Mean-variance Estimation Networks:}
Mean-variance Estimators (MVEs) estimate the mean and variance of data labels, where the variance is the square of the aleatoric uncertainty $\sigmaal$~\cite{2023arXiv230208875S, 2019arXiv190603260D, Seitzer2022}.
The MVE loss function is set to the $\beta$-negative log-likelihood: 
$\lossmve{} = \nll{}(\betanll{})$, 
where $\betanll$ is a hyperparameter. 
For the NLL loss, the gradient becomes small for high-variance data points, causing them to be undersampled. 
The $\beta$-NLL approach resolves this by multiplying a variance-re-weighting term $\sigma^{2\beta_\textrm{NLL}}$~\cite{Seitzer2022}. 
For $\betanll = 1$, the gradient is equivalent to that for the mean-squared error (MSE) loss.
For $\betanll = 0$, the original NLL loss is recovered.

\textbf{Unsupervised Domain Adaptation (UDA):} 
In UDA, the \datasource{} data have labels, while the \datatarget{} data do not have labels.
Common UDA approaches include adversarial methods~\cite{LCWJ2015, GBB2011, Ganin2015, Ganin2016}\ and distance-based methods ~\cite{Gretton2012, KL1951, Sun2016,  wilson_survey_2020}.
We use the distance-based method, Maximum Mean Discrepancy (MMD), wherein the loss $\lossuda{}$ is a multi-dimensional distance between latent embeddings of the source and \datatarget{} data sets~\cite{Gretton2012, swierc_domain_2023}.
In minimizing the MMD loss, these embeddings of the source and target data become aligned and include domain-invariant features, which allows the model to perform well on domain-shifted data.

\figurelatent{
    \textbf{(a)}: Example lensing images in the source domain (top) and the target domain (bottom) in bands $g$, $r$, and $z$.
    \textbf{(b)}: Isomaps of the latent space embeddings when the \modelmve{} model (left) and the \modelmveuda{} model (right) are applied to the source (triplet) and target (circle) domain data. 
 }{\label{fig:latent}}

\textbf{Combining MVE and UDA:}
We combine these two methods via their loss functions.
First, the \datasource{} and \datatarget{}  data are both passed through convolutional layers. 
UDA loss is then calculated on the source and target domain latent embedding --- i.e., the layer where extracted features are flattened into one dimension.
Then, the source domain embedding is passed into dense layers, and the MVE loss is calculated on the \datasource{} data only.
The total loss is 
$\losstot{} = \nll{}(\betanll{}) + \lossfactoruda{} * \lossuda{}$,
where $\lossfactoruda{}$ determines the weight of the UDA loss relative to the MVE loss.
The total loss is used to update all weights.

\section{Experiments}
\label{sec:experiments}

\textbf{Data:}
We use the \deeplenstronomy~\cite{deeplenstronomy, lenstronomy:2018, lenstronomy:2021} to simulate ground-based telescope images of galaxy-scale strong lenses.
Images have a pixel scale $0.263$\myarcsec{}/pixel, matching the Dark Energy Survey (DES)~\cite{Abbott2018}.
The lens galaxy light profile (\sersic{}) is assumed to be centered on the lensing mass.
We use theoretically and empirically inspired uniform priors for typical strong lensing parameter distributions.
For the lens mass, we use SIE profile, Einstein radius $\einsteinrad{}$, eccentricity $\{\eccentricity{l}{1}, \eccentricity{l}{2}\}$, and external shear $\{\gamma_{1},\gamma_{2}\}$~\cite{Narayan1996}. Two-dimensional source eccentricity is $\{\eccentricity{s}{1}, \eccentricity{s}{2}\}$.
For the lens and source light profiles, we use \sersic{} profiles with distribution index $n$, and scale radius $R$. 
The relative angular positions between the background and lens galaxies are $\{x, y\}$. 
Prior ranges for all simulation parameters can be found in \mytable~\ref{table:setup}.
\tablesetup{
Prior distributions of the simulation parameters for training and test sets.}{\label{table:setup}}

We use three photometric bands ($g$, $r$, $z$) to get rich image morphologies during training.
To generate realistic galaxy colors, each simulated lens galaxy is assigned a redshift in the range $z_l < 0.7$, and each background galaxy a redshift in the range $1.27< z_s < 2$ according to the DES Y3 Gold Catalog ~\cite{SN2021, zaborowski2023}.
Each galaxy is randomly assigned a color from a real galaxy according to the assigned redshift~\cite{Dey2019}.
So that each lens galaxy is visible but not saturated, we use a lower limit on the apparent magnitude for all bands ($\{m_g, m_r, m_z\} > 17.5$) and an upper limit for any one of the bands ($\{m_g, m_r, m_z\} < 21$). 
For the background galaxy, we use the limits $\{m_g, m_r, m_z\} > 17.5$ and $m_g < 22$~\cite{zaborowski2023}. 
Redshifts are used solely for colors and are independent of the lensing configuration.

We induce a domain shift between the source and the target domains in terms of image noise. 
The \datasource{} data has noise characteristics that represent a nearly noiseless image: read noise is 0 e$^-$; no sky brightness is added; the exposure time is 1000 seconds (set high to minimize Poisson/shot noise); the number of exposures is 10; the zero-point magnitude is 30; the CCD gain is 6.083 e$^-$/count; seeing is $0.9$\myarcsec{}\ (moderate for modern optical cosmic surveys)~\cite{Abbott2018, GB2015, deeplenstronomy}.
In contrast, the \datatarget{} data has noise that mimics DES: the read noise is 7.0 e$^-$, the exposure time is 90 seconds (typical of modern optical cosmic surveys)~\cite{Abbott2018, Dey2019}, and the number of exposures, the magnitude zero point, the sky brightness, and the seeing are sampled from empirical distributions~\cite{Abbott2018}.
Our dataset has 100,000 objects each for the \datasource{} and \datatarget{}  data. 
We use a 70/10/20 (training/validation/test) split for all data.
The test set is used for all results in this paper.
All images have a shape of 40$\times$40 pixels.
The dataset uses $\sim9$ GB.
Project data can be found on \href{https://zenodo.org/records/13647416}{Zenodo}.
An example image is shown in \myfig{}~\ref{fig:latent}(a).

\textbf{Model Optimization:}
We build our models using \pytorch{}~\cite{pytorch}. 
The network has three convolution blocks (each with a convolution, maxpooling, and batch normalization layer) followed by three dense layers with 128, 32, and two nodes, respectively. 
MVE techniques present challenges for training --- e.g., highly fluctuating loss functions~\cite{Seitzer2022}.
We found that a batch size of 128, a learning rate of 0.001, and default settings for \texttt{AdamW} provided optimal model performance~\cite{loshchilov2019}.
We considered scheduling of the hyperparameters for the UDA and MVE losses.
We found the best results with $\betanll=0.5$~\cite{Seitzer2022} and  $\alpha_{\mathrm{UDA}}=1.4$. 
Over 150 epochs of training, we selected the best model as the one that minimized the MVE loss on \datasource{} data.
For some seeds, the \modelmveuda{} model pathologically predicts a mean or variance of zero and does not recover --- further investigation of this is out of scope.
\myappendix~\ref{sec:app:network} briefly discusses architecture and training details.
Project code can be found on \href{https://github.com/deepskies/DomainAdaptiveMVEforLensModeling}{Github}.
\figureresidual{
    \textbf{(a)}: The four left plots show the residuals of the Einstein radius inference for the \modelmve{} model on the \datasource{} data (purple, solid),  the \modelmve{} model on the \datatarget{} data (pink, dashed), the \modelmveuda{} model on the \datasource{} data (orange, solid), and the \modelmveuda{} model on the \datatarget{} data (cyan, dashed).
    The points and error bars are the residuals from the means and the aleatoric uncertainties for randomly selected objects from the test set in each domain.
    The sample standard deviation is shaded with the corresponding color for each plot.
    The fifth (right) plot shows the binned average aleatoric uncertainty $\sigmaalmeanbin$.
    \textbf{(b)}:  Uncertainty coverage on the Einstein radius for the \modelmve{} and \modelmveuda{} models applied to \datasource{} and \datatarget{} data for five randomly seeded models. 
    The bold lines highlight the Selected model. 
    Panels \textbf{(a)} and \textbf{(b)} share the same colors and line styles. 
}{\label{fig:residual}}

\section{Results: UDA improves MVE performance on the target domain}
\label{sec:results}

\tableresultsmodel{
    Mean residual $\residualmean$ and mean aleatoric uncertainty $\sigmaalmean$ of the for the ``Selected'' Model; 
    the ``Median'' $\residualmedian{}$ and $\sigmaalmedian{}$ across five \modelmve{} and five \modelmveuda{} model fits. 
    The units are arcsec (\myarcsec{}).
    Calculations are described in \mysection\ref{sec:results}.
    \myappendix~\ref{sec:app:stability} briefly discusses quantities for the four other models.
}{\label{table:performance}}

We trained five models that differed in their weight initialization. 
The median results across initializations are consistent with the ``Selected'' model (\mytable~\ref{table:performance}). 
Therefore, unless otherwise stated, we refer only to results of the ``Selected'' model for clarity of presentation.
For the mean residual $\residualmean{}=\langle \residualbasicdefinition{} \rangle$ and aleatoric uncertainty $\sigmaalmean{}$, we take the mean over all the data for a single model.
Ideally, the successful combination of MVE and UDA  (\modelmveuda) would perform comparably to the \modelmve{} model on the \datasource{} data.
When applied to \datasource{} data, the \modelmveuda{} model has a higher mean residual $\residualmean{}$ by $\sim0.015$\myarcsec{} compared to the \modelmve{} model (\mytable~\ref{table:performance}(a), \myfig{}~\ref{fig:residual}(a; four left plots)). 
The mean uncertainty of the \modelmveuda{} model is approximately twice that of the \modelmve{} model (\mytable~\ref{table:performance}(b), \myfig{}~\ref{fig:residual}(a; fifth, right plot)). 
At the same time, the \modelmveuda{} model is better calibrated (less underconfident) than the \modelmve{} model (\myfig{}~\ref{fig:residual}(b)). 
For \datatarget{} data, however, the \modelmve{} model has a high mean residual $\residualmean{}=0.0818$\myarcsec{}, twice the \modelmveuda{} model's mean residual $\residualmean{} = 0.0425$\myarcsec{} (\mytable~\ref{table:performance}(a)).
In contrast, the \modelmve{} model has a low mean uncertainty $\sigmaalmean{} = 0.0253$\myarcsec{}, half the \modelmveuda{} model's mean uncertainty $\sigmaalmean{}=0.0503$\myarcsec{} (\mytable~\ref{table:performance}(b)).
Commensurately, the \modelmve{} model is significantly overconfident, while the \modelmveuda{} model is only slightly overconfident (\myfig{}~\ref{fig:residual}(b)) on target data. 

The \modelmveuda{} model uncertainty is higher at both low and high values of $\einsteinrad{}$ (\myfig{}~\ref{fig:residual}(a) and \myfig{}~\ref{fig:residual}(a; fifth, right plot))).
The high uncertainty for the \modelmveuda{} model at low $\einsteinrad{}$ may be due to low image resolution or high seeing, such that smaller lensing arcs could be obscured.
The high uncertainty at high $\einsteinrad{}$ may be caused by the image being too small to contain the lensing arcs.
The residuals and uncertainties for both models are slightly larger than uncertainties assumed in some studies $\sim0.01$\myarcsec{}~\cite{10.1093/mnras/stad3514} but comparable to those from traditional modeling techniques $\sim 1$-$5\%$~\cite{2011ApJ...727...96R, 2013ApJ...777...97S}.
In \myfig{}~\ref{fig:latent}(b), we find that the \datatarget{} and \datasource{} embeddings do not overlap for the \modelmve{} model.
In contrast, the embeddings overlap almost completely for the \modelmveuda{} model, and the points exhibit a gradient in the Einstein radius.
These items indicate that the embedding vectors of both are correlated with $\einsteinrad{}$, but only the \modelmveuda{} embedding has accurate alignment across domains. 
Lastly, the coverage of the \modelmve{} model varies significantly on the \datatarget{} data across initializations, but performance is stable for \modelmveuda{} (\myfig{}~\ref{fig:residual}(b)).
We find DA is essential to MVE for better calibrated, consistent, and accurate performance on domain-shifted datasets.

\section{Summary and Outlook}
\label{sec:conclusion}

In this work, we provide the first demonstration that unsupervised domain adaptation (UDA) significantly improves the performance of mean-variance estimator (MVE) models on unlabeled \datatarget{} data.
We predicted the Einstein radius of strong gravitational lenses with MVEs (\mysection\ref{sec:methods}).
We incurred a domain shift between the \datasource{} and \datatarget{} domains so that the \datasource{} images are approximately noiseless, and the target images have noise characteristics similar to DES (\mysection\ref{sec:experiments}). 
When applied to the noisy \datatarget{} data, the \modelmveuda{} model is significantly better calibrated, more consistent across weight initialization, and more accurate than the \modelmve{} model (\myfig{}~\ref{fig:residual}(a) and \mytable~\ref{table:performance}(c,d)). 
Similar approaches may improve neural network model performance when applied to real, observational data.

\bibliography{neurips_custom_2024, neurips_general_2024, neurips_software_2024}
\bibliographystyle{plain}

\newpage
\appendix
\onecolumn

\begin{ack} 
\section{Funding}
\label{sec:app:funding}

\ackfundingdeepskies
\ackfundingdoegeneral
\ackfundingdoeecanord

\section{Author Contributions}
\label{sec:app:contributions}

\contribagarwal
\contribciprijanovic
\contribnord

We thank the following colleagues for their insights and discussions during the development of this work: Rebecca Nevin.

\end{ack}

\section{Software Attribution}
\label{sec:app:software}

We used the following software packages:
\astropy{}~\cite{astropy:2013,astropy:2018, astropy:2022},
\deeplenstronomy{}~\cite{deeplenstronomy},
\lenstronomy{}~\cite{lenstronomy:2018, lenstronomy:2021},
\matplotlib{}~\cite{matplotlib},
\numpy{}~\cite{numpy},
\pandas{}~\cite{pandas}
\python{}~\cite{python},
\pytorch{}~\cite{pytorch},
\scipy{}~\cite{scipy:2001, scipy:2010},
\seaborn{}~\cite{seaborn},
\sklearn{}~\cite{sklearn_api, scikit-learn},
\torch{}~\cite{torch},
\torchvision{}~\cite{torchvision2016},


\section{MVE Network Architecture}
\label{sec:app:network}

\tablenetworkmve{
    The architecture of the MVE network. 
    The first column lists the layer type, the second lists the dimensionality of the output from that layer, and the third column lists the parameters of that layer; $k$ is the kernel size, and $s$ is the stride. 
    The final layer outputs the mean and variance. 
}{\label{table:networkmve}}

See \mytable~\ref{table:networkmve} for the detailed MVE network architecture.
There are 112,866 trainable parameters. 
We note that the activation function for the final dense layers is chosen to be sigmoid rather than ReLU, since ReLU predicts a value of zero for any negative input, encouraging predictions of zero mean or variance. 
This issue can also be solved by alternative approaches, such as the use of Leaky ReLU or other activation functions that disincentivize a prediction of zero.

\section{Model inference with varied weight initializations}
\label{sec:app:stability}

We performed experiments five times, each with a different random seed for the network weight initialization.
All models received the same optimization procedure (\mysection\ref{sec:experiments}).
The performance of \modelmve{} model on the target data sets is inconsistent across the seed choices. 
In contrast, the \modelmveuda{} model performs consistently slightly worse than the \modelmve{} model on the \datasource{} data across varied weight initialization. 
However, unlike the \modelmve{} model, it consistently performs equally well on the \datatarget{} data as on the \datasource{} data.
This indicates UDA adds stability against the domain shift and is necessary for the application of MVE to datasets with a domain shift. 
For some initializations, the \modelmveuda{} model training starts with predictions of zero for the mean or variance, which is erroneous.
Further training does not improve the performance. 
Investigating this pattern in detail is outside the scope of this work.
We chose seeds where this pathological behavior does not occur in the first epoch of training. 


\tableresultsmodelappendix{
    Mean residual $\residualmean{}$, mean aleatoric uncertainty $\sigmaalmean{}$, mean correlation coefficient $\correlationmean{}$, and mean NLL loss $\nllmean$ across each data set for each model, \modelmve{}, \modelmveuda{}.
}{\label{table:performance_appendix}}

\section{Computational costs for experiments}
\label{sec:app:computationcosts}

All computing was executed on an NVIDIA A100 GPU with 40GB memory. 
These computations were performed on the Fermilab Elastic Analysis Facility~\cite[EAF;][]{eaf}.
Training with and without UDA require the same amount of time, $\sim2.5$ hours.

\end{document}